\documentclass{article}
\usepackage{sao2,psfig}
\issue{2001, 51, 140-150}
\begin{document}
\title{Velocity dispersion of stars and gas motion in double-barred galaxies
\thanks{Based on observations collected with the 6m telescope of the Special
Astrophysical Observatory (SAO) of the Russian Academy of Sciences (RAS),
operated under the financial support of the Science Department of Russia
(registration number 01-43}
}
\author{A.V. Moiseev}
\institute{\saoname}

\maketitle

\begin{abstract}

The current state of the problem  of double-barred galaxies investigation
is reviewed. The necessity for application of the panoramic spectroscopy
methods to a detailed study of kinematics of these objects is being proved.
The first results of observing double-barred galaxies at the 6m telescope
using the multipupil spectrograph MPFS and the scanning interferometer
Fabry-Perot are described.
\keywords{galaxies: kinematics and dynamics --- galaxies: spiral --- galaxies:
structure}
\end{abstract}

\section{Introduction}

According to the latest statistical estimates (Selwood \& Wilkinson, 1993;
Knapen, 1999; Knapen et al., 2000b; Laine et al., 2001), galaxies with central bars account for
50--70\% of the total number of nearby disk galaxies. Since the distribution
of the gravitational potential in the region of the bar is not
axisymmetric (one generally says of a triaxial potential), then the motion of stars
and gas clouds is  different from circular. This fact is confirmed by
both direct observations of gas velocity fields in barred galaxies (see,
for instance, Afanasiev \& Shapovalova, 1981; Duval \& Monnet, 1985; Knapen
et al., 2000a) and numerous model calculations, (see, for example
Athanassoula, 1992; Combes, 1994; Lindblad, 1996; Vauterin \& Dejonghe, 1997).
By now it
can be considered to be proved that the bars are dynamically decoupled subsystems,
where the motions of gas and stars being of different kind. Inside the
bar there exist several ``families'' of stable periodic orbits that form
an off-beat ``backborn'' of the bar (Contopouls \& Grosbol, 1989). The main
orbits that maintain the shape of the bar are elongated parallel to the bar
and belong to family $x_1$. If two Inner Lindblad Resonances (hereafter ILR)
are present in the galaxy then stable  orbits of family $x_2$ are present
between them, which are perpendicular to the main bar. The location of the
resonances is defined by the relationship between angular velocity of
differential rotation of the galaxy and angular velocity of  the bar rotation.

Such motions are possible only in a collision-free stellar subsystems. The gas
subsystem (where by the gas particles are implied individual clouds of
interstellar gas) is collision. Therefore the stable existence of intersecting flows
is impossible in it. Gas clouds follow ``smoother'' paths. The flows of
gas onto a relatively slow rotating bar, which occurs at a velocity of
50--100 km/s, leads to formation of strong shock fronts at the leading edges
of the bar (Athanassoula, 1992).

Numerous model calculations show that the bar takes up effectively angular
moment from the gas of the disk, which results in formation of gas streams
towards the galaxy centre. Although a detailed hydrodynamic modelling (Levy
et al., 1996) shows that apart from the flows towards the centre (inflows) there
exists a reverse motion (outflow) caused by the saddle point of the
gravitational potential.  However, it should be noted that the bar promotes
increasing of gas concentration in the circumnuclear region. This is confirmed
by the real observations. Sakamoto et al. (1999) showed that the molecular
gas concentration inside the central kiloparsec in barred galaxies is by
an order of magnitude higher than that in galaxies without bars. The bar is
often treated as one of the basic mechanisms of transport of the interstellar
gas to the circumnuclear region where it becomes a fuel for an active
nucleus or a circumnuclear burst of star formation.

In the relation between the bar and the phenomenon of active
(Seyfert) nucleus, two points should be noted. Firstly,
according to the latest estimates of Knappen et al. (2000b), the relative
proportion of bars in Seyfert galaxies is negligibly larger than one
in galaxies of the control sample  (80\%$\pm 8$\% and 60\%$\pm9$\%,
respectively). Also Laine et al. (2001) provide  more contrast values
of the bar percentage (73\%$\pm 6$\% for Sy and 50\%$\pm7$\%
for non-Sy galaxies).  Knappen et al. and Laine et al.  performed  measurements
using the data of surface photometry in the near IR range where the influence
of dust is negligible, and the contribution of the old stellar population into
luminosity is significant, which facilitates a more reliable detection of
bars as compared to optical observations.

Secondly, the bar transports gas not to the  centre, but to the ILR
region where the gas is concentrated in a ring of radius of a few hundred
parsec. Thus, there is a problem of taking up the angular rotation moment from
the gas situated at a distance of 100--1000 pc from the centre and
transportation of it to the region of effective gravitational forces of the
central supermassive black hole (at a distance of $<1-10$ pc). In particular
detail, this problem is discussed in the survey  ``Fueling the AGN'' (Combes, 2000)

\section{Double-barred galaxies}
\subsection{The historical review}

One of the refined performances of the task of matter transfer to the active
nucleus was suggested by Shlosman et al. (1989). They showed that in the gas
disk (ring) of radius of a few hundred pc (which is formed inside a
large-scale bar), a new bar can be formed under the action of a bar-forming
instability. And this bar again forms flows of gas towards the nucleus. The
characteristic scale on which gas is concentrated  is less than
0.1 $R_{bar}$, where $R_{bar}$ is the bar radius. For this reason, the system
of two bars is capable of ``sweeping'' the interstellar medium on scales of
a few kiloparsec and concentrate it at distances of 1--10 pc from the centre.
The process of angular moment transfer will further be defined by the turbulent
viscosity of the accretion disk formed around the central supermassive object.

At nearly the same time with the paper by Shlosman et al. (1989), in which it
was spoken only of a purely gaseous inner bar, Pfenninger \&
Norman (1990), using a numerical modeling, demonstrated the formation of the
second bar as a result of development of instability in the stellar
self-gravitating disk. Of particular interest is the conclusion that the internal
bar does not necessarily  rotate at the same angular velocity as the outer
one does. The most stable is the configuration in which the corotation
radius of the inner bar coincides with the position of the ILR of the main bar.
Such an elaborate numerical modeling  of the stellar-gaseous disks was
carried out  by Friedli \& Martinet (1993), in which a double bar was
treated as one of the stages of evolution of barred galaxies. In 2001 year new
hydrodynamical simulations of the gas behaviour in double bars are
presented by Heller et al.(2001), Maciejewski et al. (2001) and Shlosman \& Heller (2001).

And what do the observations show? As early as in 1975, de Vaucouleurs found
that an inner (stellar) circumnuclear bar (turned
by $30\degr$ with respect to the outer one) was sharply decoupled on the
optical images of NGC\,1291 (de Vaucouleurs, 1975). The same
galaxy was investigated by Jarvis et al. (1988) and they gave one more example of
a double-barred galaxy, NGC\,1543. However, the first attempt of systematic
observational studies of this phenomenon was made by Buta \& Crocker (1993) who published a list of 13
galaxies with ``inner circumnuclear bars''.

The observed position angle between the inner and outer bars in these galaxies
varies over a wide range, from $20\degr$ to $90\degr$. Since more than half
of galaxies have a small inclination with respect to the line of sight,
$i<30\degr$, then a projection must have a minor effect on the estimate of the real angle between
the bars in the galaxy plane. Therefore, in the opinion of Buta \& Crocker
(1993), the second bar can be oriented arbitrarily relative to the outer one.
This question will be discussed in more detail in the next Section 2.2.

Wozniak et al. (1995) suggested a method of searching and classification of
double bars, which is based on the isophote analysis of galactic images.
By the turn of the position angle (PA) and by the change of ellipticity  of
the inner isophotes, the galaxies were divided into one-bar objects, double-bar ones,
and containing a bar with a triaxial bulge. The latter term was applied
to galaxies that demonstrate a smooth variation of the PA in contrast to a
``pure'' bar where the $\rm PA\approx const$. By applying this technique to
the optical CCD images of 36
barred galaxies, Wozniak et al. (1995) found  an ``inner triaxial
structure'' in 2/3 of the cases, i.e. a second bar or a triaxial bulge. The
galaxies of their sample (Friedli et al., 1996) are observed in the near IR
range.
 Double bars were confirmed in all (excluding  two, NGC\,6951 and NGC\,7479) objects.
 Similar observations made
over the last few years in the IR range (where the influence of dust and
young stars is small as compared with the optical range) have extended
substantially the list of galaxies suspected to have double bars (Jungwiert
et al., 1997; Greusard et al., 2000). An image of the double-bar galaxy
NGC\,2950 we have obtained with the 6m telescope is displayed in Fig.\,1
for example.

Table 1 gives the list of all similar objects that we have found in references.
The first column lists the name of the galaxy; the second column indicates the
morphological type from the catalogue RC3; $a_s$ and $a_p$ are the sizes
of the semiaxes of the inner and 

\clearpage
\begin{onecolumn}
\begin{center}
\topcaption{
A list of double-barred galaxies}
\tablehead{\hline
\multicolumn{1}{l}{Name}&\multicolumn{1}{l}{Type}
 &\multicolumn{1}{l}{$a_s$}&\multicolumn{1}{l}{$a_p$} &\multicolumn{1}{l}{AGN}&
\multicolumn{1}{l}{References}&\multicolumn{1}{l}{ Notes}   \\
\hline}
\tabletail{\hline}
\begin{supertabular}{llrrrll}
ESO 437-67& SBab &  3&  32&      & (15)& ? \\
ESO 508-78& SBa  &   &    &      & (7) &   \\
ESO 215-31& SBb  & 10&  47&      & (13)& \\
ESO 320-30& SABa &  5&  37&      & (13)& \\ 
ESO 443-17& SB0/a&  6&  15&      & (13)& \\
Mrk  573  & SAB0 &  5&  20&  Sy2 & (1), (3), (19) & \\
Mrk 1066  & SB0  &  3&  16&  Sy2 & (1) &  \\
NGC  470  & SAb  &  8&  32&      & (12), (24) & $B+T$    \\
NGC  613  & SBbc &  5&  59&  Sy  & (15) &          \\
NGC 1079  & SABa & 17&  32&      & (15) &  $T+B$\\ 
NGC 1097  & SBb  & 10&  80&  Sy1 & (7),(11),(12),(20),(24) &  $B+T$ \\
NGC 1291  & SB0/a&   &    &      & (7), (14), (23) & \\
NGC 1300  & SBbc &   &    &      & (4)&  $T+B$ \\
NGC 1317  &SAB0/a& 11&  50&      & (17), (22) & \\
NGC 1326  & SB0/a& 10&  50& LINER& (7), (24) & \\
NGC 1353  & SAbc &  4&  14& LINER& (15)  & ? \\
NGC 1365  & SBb &  8& 150& Sy1.8& (4),(15)  &  $T+B$  \\
NGC 1371  & SABa & 10&  60&      & (24) & \\
NGC 1398  & SBab & 14&  36&  Sy  & (15) &  $T+B$ \\
NGC 1433  & SBab &  5& 100&  Sy2 & (4),(5),(15),(24) & \\
NGC 1512  & SBab &  6& 150&      & (15) &    $T+B$ \\
NGC 1543  & SB0  &   &    &      & (7),(14)& \\
NGC 1566  & SABbc&   &    &  Sy1 & (4) &     $T+B$ \\
NGC 1672  & SBbc &   &    & Sy2  & (4)   &  $T+B$ \\
NGC 1808  & SABb &  3&  ? & Sy2 & (15) & \\
NGC 2217  & SB0/a&  8&  37&      &(4), (15) & \\
NGC 2273  & SBa  &  8&  24&  Sy2 &  (2), (17) & \\
NGC 2442  & SABbc&   &    &      & (4) &  $T+B$ \\
NGC 2681  &SAB0/a&  5&  29& LINER& (9), (24)&  $3\beta$  \\
NGC 2859  & SB0  & 12&  46&      & (8),(24) &       \\
NGC 2880  & SB0  &  &   &      & (8) & ?     \\
NGC 2935  &SABb  & 11&  25&      & (15) &   $T+B$? \\
NGC 2950  & SB0  &  6&  38&      & (12),(24) &    \\
NGC 3081  &SAB0/a& 10&  37&   Sy2& (3),(6),(7),(12),(17),(24) & \\
NGC 3185  & SBa  &  2&  30&   Sy2& (1),  (8) &  ?      \\
NGC 3275  & SBa  &  5&  34&      & (17)& \\
NGC 3358  & SABab&   &    &      & (7) & \\
NGC 3368  & SABab&  4&  24& LINER& (15)&  $3B$ \\
NGC 3393  & SBab &  2&  13&   Sy2& (3), (13), (15) & \\
NGC 3412  & SB0  &  &   &      & (8)&   ?\\
NGC 3516  & SBO  &  6&  22&   Sy1& (16) & \\
NGC 3786  & SABa &  7&  25& Sy1.8& (1)&            \\
NGC 3941  & SB0  &  &   & Sy2  & (8)&   ?   \\
NGC 3945  & SB0  & 20&  42& LINER& (9),(24) &   $3B$ \\
NGC 4262  & SB0  & 10&  14&      &  (21)&             \\
NGC 4274  & SBab & 10&  39&      &  (21)&             \\
NGC 4314  & SBa  &  6&  75& LINER& (8), (21)&  \\
NGC 4321  & SABbc& 10&  66&      &  (18), (21)&       \\
NGC 4340  & SB0  &  5&  52&      & (12), (24)&       \\
NGC 4371  & SB0  & 24&  43&      & (21), (24) &  $3B$     \\
NGC 4593  & SBb  &  2&  60&   Sy1& (24) &  $B+T$      \\
NGC 4594  & SAa  & 10&  68& LINER& (10) &   \\
NGC 4643  & SB0/a& 17&  51& LINER&  (8),(21)&   \\
NGC 4736  & SAab & 10&  26& LINER& (20) &            \\
NGC 4754  & SB0  &  7&  21&      &  (21) &               \\
NGC 4984  & SAB0 &  4&  30&      & (15) & \\
NGC 5101  & SB0/a&  2&  50& LINER& (15) & \\
NGC 5365  & SB0  & 18&  33&      & (17) & \\
NGC 5566  & SBab &  6&  24& LINER&  (15)&      \\
NGC 5728  & SABa &  4&  44&  Sy2 & (7),(20),(24) & \\
NGC 5850  & SBb  &  9&  84&      & (7),(24)& \\
NGC 5905  & SBb  &  6&  37&      & (12), (24)&       \\
NGC 6300  & SBb  &  4&  44&   Sy2& (17) &?  \\
NGC 6782  & SB0/a&  3&  26&      &  (7),(12),(15),(24) & \\
NGC 6951  & SABbc&  5&  44&  Sy2 &  (12), (17) & ?   \\
NGC 7007  & SA0  &  4&   9&      & (17) & \\
NGC 7098  & SABa & 14&  57&      & (12), (7), (24)&   \\
NGC 7187  & SAB0 &  9&  28&      & (24) &   $3B$           \\
NGC 7479  & SBc  &  ?&  46& LINER& (4), (12)& ?  \\
NGC 7702  & SA0  & 10&   ?&      & (24) &  $T+B$ \\
NGC 7743  & SB0  & 10&  57&  Sy2 & (24) &  $B+T$   \\
\end{supertabular}
\end{center}
\end{onecolumn}
\begin{table}[h]
\caption{References to Table 1}
\begin{tabular}{rll}
\hline
(1)   &    VRI                &  Afanasiev et al. (1998a) \\
(2)   &    VRI                &  Afanasiev et al. (1998b) \\
(3)   &    JHKL$'$            &  Alongso-Herrero et al. (1998) \\
(4)   &    plates & Baumgart \& Peterson (1986)\\
(5)   &   plates & Buta (1986) \\
(6)   &   I                   & Buta (1990) \\
(7)   &   BVI                 & Buta \& Crocker (1993) \\
(8)   &   BR                  & Erwin \& Sparke (1998)  \\
(9)   &   HST                 & Erwin \& Sparke (1999) \\
(10)  &    ---              &  Emsellem \& Ferruit (2000)\\
(11)  &    K                  &  Forbes et al. (1992) \\
(12)  &    JHK                &  Friedli et al. (1996) \\
(13)  &    JK$'$              &  Greusard et al. (2000) \\
(14)  &    gr                 &  Jarvis et al. (1988)   \\
(15)  &    JHK                &  Jungwiert et al. (1997) \\
(16)  &    V                  &  Moiseev et al. (2000) \\
(17)  &    K                  &  Mulchaey et al. (1997) \\
(18)  &    I                  &  Pierce (1986) \\
(19)  &    HST                &  Pogge \& De Robertis (1995)\\
(20)  &    JHK                &  Shaw et al. (1993) \\
(21)  &    JHK                &  Shaw et al. (1995) \\
(22)  &    plates             &  Schweizer  (1980) \\
(23)  &    plates             &  de Vaucouleurs (1975) \\
(24)  &    BVRI,$H_\alpha$    &  Wozniak et al. (1995) \\
\hline
\end{tabular}
\end{table}

\vspace{0.5cm}
outer bar in arcseconds,
respectively\footnote{Hereafter indices ``p'' and ``s'' mark values which
relate to the primary and secondary bars, respectively}
 (if
these data were given by the authors). In the next columns are shown the type
of nucleus activity (from the NED database) and the references. The symbol
? in the last column marks the uncertain data. $B+T$ denote the combination of the triaxial bar
and the bulge; $3B$ is the ``triple bar''. In Table 2 is presented the list
of literature sources with indication of the observational method (photographic,
CCD or IR photometry in the bands). The case of NGC\,4594 is distinguished.

This
is a known nearby edge-on galaxy ``Sombrero'' in which Emsellem \& Ferruit
(2000) found as many as two bars on the basis of 2D spectroscopy with
the integral field spectrograph TIGER.

Attention is attracted by the fact that 30\% of all the galaxies in Table 1
are Seyfert, and 15\% of LINER-type galaxies. However, such a number of active
objects (an order of magnitude larger!) relative to the  field galaxies
is most likely caused by strong selection effects. Really, in the paper by
Mulchaey \& Regan (1997) no noticeable discrepancy in the relative number  of
double bars in the samples of Seyfert and normal (non-active) galaxies has been found;
see also the discussion  in Friedli (1999) and in Laine et al. (2001).
The influence of
selection effects can also explain the relatively large percentage (56\%) of
early-type (S0--Sa) galaxies because the second bar in such galaxies can be
easier defined. So, Erwin \& Sparke (1998)  argue that in no less than 20\%
of S0--Sa barred galaxies the second bar is observed as well.

\begin{figure*}
\centerline{\psfig{figure=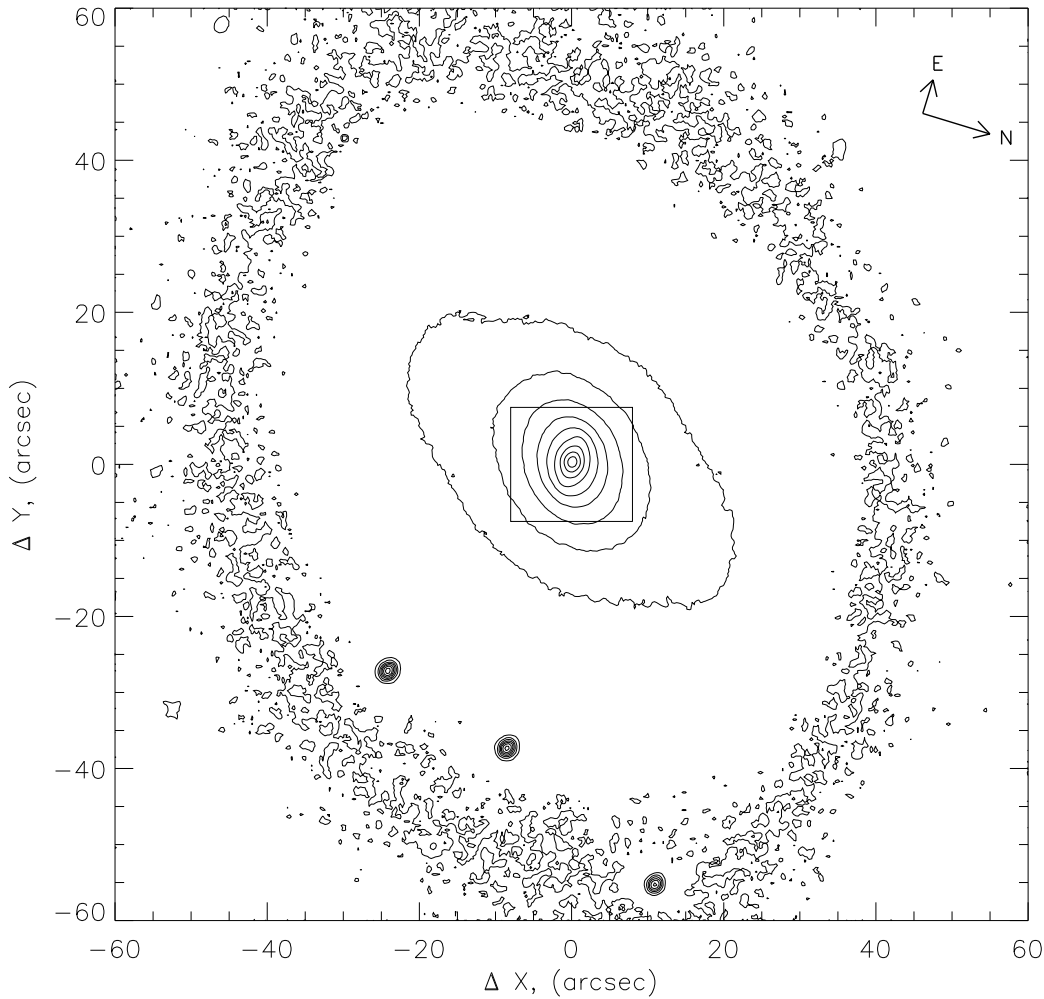,width=14cm}}
\caption{The NGC\,2950 image obtained with the BTA focal reducer SCORPIO
in the medium-band filter centered on
 $\lambda=7550\AA$. The region of the secondary bar observed with the
multipupil spectrograph (Fig. 2) is marked with the box.}
\end{figure*}

Let us make some remarks concerning the terms
 used. Double-barred galaxies are termed also
``bar-within-bar'' systems. The outer bar is called ``primary'' while the inner
one is named ``secondary''.

\subsection{Relative orientation of bars}

In the above mentioned paper Friedli \& Martinet (1993) showed that if the
two bars have the same angular velocity ($\Omega_p=\Omega_s$), only two stable
configurations are then possible --- the bars are parallel or perpendicular
to one another. The former case is a trivial variation of radial
distribution of surface brightness along the bar axis.
The case of the
perpendicular bars is more intriguing, however, it is difficult
to confirm  by observations, because only the position angle between the
projections of the bars in the sky plane can actually be measured, and one
has to introduce a correction for the inclination of the galaxy plane to the
line of sight ($i$). The task is not  trivial because the point in question
is obviously the projection of three-dimensional structures. It is not always
possible to unambiguously determine $i$ from the ellipticity of the outer
isophotes, because in the outer parts of barred galaxies rings are often observed,
elliptical in the galaxy plane. Sometimes, the model of the mutually perpendicular
bars describes well the observed surface brightness distribution (see, for
instance, the paper of Moiseev (1998) about the 2D-decomposition of Mrk\,573).
However, in most of the galaxies observed nearly face-on, the angle between
the bars differs substantially from $90\degr$ (Buta \& Crocker, 1993). In the
papers by Friedli \& Martinet (1993), Friedli et al. (1996), Jungwiert
et al. (1997), which were already discussed above, attempts were made to
take account of the effect of projection of the bars onto the sky plane and,
first of all, to test the hypothesis of mutually perpendicular bars. And
all these authors came to the conclusion that there is no distinct characteristic
value of the angle between the bars.

In the existing models this refers  to the case of a dynamically independent
inner bar. It is usually believed
that from general considerations $\Omega_s>\Omega_p$ (but Heller et al. (2001)
studied the secondary gaseous bar with   $\Omega_s<\Omega_p$).

Since the stellar component
in galaxies is collisionless, then this situation (a bar rotates inside a bar)
is possible and is even reproduced in numerical experiments of Friedli \& Martinet
(1993). Maciejewski \& Sparke (2000) showed that in a system of two dynamically
independent bars, rotating one inside the other, there exist several families of
orbital loops maintaining this configuration (similar to orbits $x_1$ and $x_2$
in a galaxy with a single bar). In accordance with the opinion of these and
other authors, a configuration, in which the relation between angular velocities is not
arbitrarily but such that the corotation radius of the inner bar lies
near the resonance IRL of the outer bar, is stable (Pfenninger \& Norman, 1990;
Friedli \& Martinet, 1993; Maciejewski \& Sparke, 2000). If the rotation curve
of the galaxy is such that there is no ILR in the primary bar, then, according
to Maciejewski \& Sparke (2000), the secondary stellar bar cannot  to exist.

\section{Two-dimensional spectroscopy of double bars}

\subsection{Necessity of 2D-methods}
The numerous observational papers (see Table 2) point to the fact that in the
case of double bars we are probably faced with some new structural feature
of barred galaxies. However, all the papers enumerated above have a significant
disadvantage, the presence of the secondary bar can be found only from
photometric data, if on the galaxy image, some extended structure is observed
inside the primary bar. The formal application of the results of isophote
analysis, as described by Wozniak et al. (1995), allows even ``triple
bars'' to be decoupled (Erwin \& Sparke, 1998; 1999; Jungwiert et al., 1997;
Friedli et al., 1996) without any reasoning for possible dynamic behaviour
of such configurations. At the same time, the observed photometric structure
can, in our opinion, be explained in a less exotic manner, without involvement
of the secondary and the third bar. It is difficult to
separate the following possibilities  using only the data of the surface
photometry:
\begin{itemize}
\item
The dynamically independent secondary bar ($\Omega_s\ne\Omega_p$).
\item
The $x_2$-orbits (perpendicular to the primary bar) between two
inner ILR resonances of the primary bar. In contrast to the previous case,
this structure is not dynamically independent.
\item
The elliptical ring in the disk plane at the ILR resonance of the primary bar.
A good example is the galaxy NGC\,6951, the HST observations of which
showed that the ``secondary bar'' is a ring resolved into separate star
formation regions (Barth et al., 1995).
\item
The polar disk (ring) inside the primary bar. Similar structures have been
found in a number of ordinary galaxies (Sil'chenko, 2001) such as NGC\,2841
(Afanasiev \& Sil'chenko, 1999), NGC\,7280 (Afanasiev \& Sil'chenko, 2000),
 and others.
\item
The projection of the central part of an oblate bulge inside the primary
bar onto the sky plane. An illusion can be created of a ``second bar'',
the major axis of which is virtually coincident with the line of nodes of
the disk.
\item
The complex distribution of dust and star formation regions inside the
primary bar impedes the study of the circumnuclear region and can create
illusion of the presence of a second bar. This problem is solved  partly
by observations in the near IR range.
\end{itemize}

That the interpretation of photometric observations of double bars is
unambiguous is suggested by the fact that in the paper by Regan \& Mulchaey
(1999), the authors, using the colour distribution maps obtained with
the HST (including those in the near IR), interpreted the structures
observed in a number of galaxies  (Mrk\,573, Mrk\,1066, NGC\,3516) as
circumnuclear spirals but not as second bars.

It seems to us that in order to understand if the second bar is readily a
new dynamically isolated region of the galaxy, additional observational data
on the kinematics of gas and stars in these strange objects are needed.
Since the motion of matter in the regions of the bars differs markedly
from circular (see Section 1), and the density distribution is different
from axisymmetric, it is necessary to use the methods of panoramic 2D
spectroscopy which makes it possible to obtain two-dimensional kinematic
characteristics in the sky plane. In 2000 a programme was started at the
6m telescope with the goal  of construction of radial velocity fields of stars
and ionized gas and two-dimensional maps of velocity dispersion of stars in
double-barred galaxies. These data will allow first of all the following
kinematic features of bars to be revealed.

\begin{itemize}
\item
The turn of the kinematic axis (the line of maximum radial velocity gradient)
in the velocity fields of stars and gas. This is one of the easily measurable
dynamical features of bars. For more details see the references in
Moiseev \& Mustsevoi (2000) and Sil'chenko (2001).

\item
The difference in the ionized gas radial velocities measured from Balmer
($H_\alpha$, $H_\beta$) and forbidden ([OIII], [NII]) lines, which is
associated with the presence of shock fronts at the edges of bars (Afanasiev
\& Shapovalova, 1996). We note that  new models by Maciejewski et al.(2001) and
Shlosman \& Heller (2001)  show that shock fronts may be absent in the
secondary bar, but a observation tests of this fact is need.

\item
The distinction of the velocity dispersion distribution from axisymmetrical case
(Miller \& Smith, 1979; Vauterin \& Dejonghe, 1997).
\end{itemize}

The author knows but a few papers concerned with a detailed study of kinematics
of galaxies in the secondary bar region. Knapen et al. (2000a) analysed the
velocity fields of ionized and molecular gas in NGC\,4321 and obtained
that  only one bar exists in this
galaxy contrary to photometric data. In a recent paper by Emsellem et al. (2001),
 data are presented on
stellar kinematics of several double-barred galaxies. In three of them
(NGC\,1097, NGC\,1808, NGC\,5728) the secondary bar region turns out to be
dynamically decoupled (the circumnuclear radial velocity peak in the sections
obtained with a ``long slit''). However, from the new data NGC\,1365 is
classified by the authors as a single bar galaxy with an inner circumnuclear
disk. In a small note of Wozniak (1999) arguments are adduced that the
``counterrotation'' of stars which is observed in cross-sections through the
circumnuclear region of NGC\,5728 may be due to the influence of the
secondary bar. It should, however, be noted that the peaks on the rotation
curves  and the ``counterrotation'' detected in one-dimensional sections are
inadequate for proving the dynamic effect of the bar and can be explained by
a number of other causes, such as merging with a companion or the
accretion of intergalactic gas (see Kuijken et al.(1996) for the discussion
of this point). The two-dimensional distributions of velocities and velocity
dispersions can give more comprehensive information.

\subsection{Observations}

From the list given in Table 1 we constructed a sample of objects to be observed
with 6m telescope BTA, basing on the following criteria:
$\delta>0^\circ$, $a_s<10''$ (the secondary bar is placed in the field of view
of the multipupil spectrograph). During the year 2000 we observed
14 objects, which makes half of all double-barred galaxies in the northern
sky. The circumnuclear regions were investigated with the multipupil spectrograph
MPFS. The field of view was $16''\times15''$, the spatial scale was $1''$/lens.
The spectral range $\lambda\lambda4900-6100$\AA\, included absorption
features (MgI, NaI, CaI, FeI) characteristic of the old (G-K) stellar population.
To construct the maps of radial velocities and velocity dispersion of stars
(further $\sigma_*$), the cross-correlation method of the spectra of galaxies
with the spectra of template stars was used. We adopted the method for working
with the 2D spectroscopy data (Moiseev, 2001). With a dispersion
of the spectrograph of 1.35 \AA/px the radial velocity measurement accuracy
was about 5--10\,km/s, the radial velocity dispersion one was 10--20 km/s for
$\sigma_*>50-70$ km/s.
The emission lines H$\beta$ and [OIII] which we employed to construct velocity
fields of ionized gas in the circumnuclear region are located within the
spectral interval. Unfortunately, because
most of the galaxies belong to old types (S0-Sa), the emission is far from
being present in all the objects.

Galaxies with bright enough emission lines were observed with the scanning
Fabry-Perot interferometer (FPI) in the 235th order (for the wavelength
6563\,\AA), the spectral resolution was about 120 km/s. Interference filter
10--15\,\AA\, wide separated the required spectral interval around the lines
$\rm H_\alpha$ or [NII]. From the observational data the  velocity
fields of ionized gas were constructed on considerably larger spatial scales
than with the MPFS because the field of view of the FPI was about $5'$. The
primary observational reduction and the wavelength scale calibration were
performed with the software developed by the author in the IDL environment.
The velocity fields were constructed with the ADHOC package
\footnote{
ADHOC  software has been written by J. BOULESTEIX (Observatoire de
Marseille).
It is free of use  http://www-obs.cnrs-mrs.fr/ADHOC/adhoc.html}.

The mean rotation curves and the radial relationships of the position angle
PA ($r$) of the kinematic axis were defined from the velocity fields by the
``tilted ring'' method (Begeman, 1989). The velocity fields were broken up
into elliptical rings of fixed width. The rotation velocity and PA values were
determined in each ring as an approximation of circular rotation. Even the
application of such a simple approximation allows one to draw a number of
conclusions as to the character of non-circular motions in the bar region
(see  Moiseev \& Mustsevoi, 2000 for discussion).

\subsection{Some results}

\begin{figure*}
\centerline{\psfig{figure=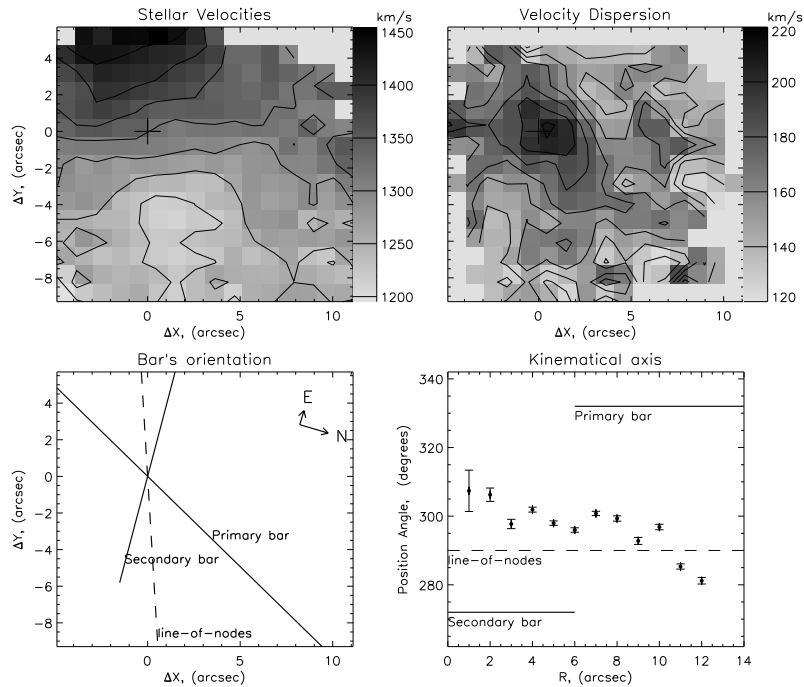,width=11cm}}
\caption{Kinematics of stars in NGC\,2950 from the MPFS observations.
(top) --- the radial velocity field and the map of the radial velocity dispersion.
The cross marks the dynamic centre. The relative orientation of the
bars in the image plane and the variations of the position angle of the
kinematic axis (``moustached'' dots) (bottom). The position angle of the bars
and the outer isophotes (the line of nodes) is obtained from the photometric
data of Friedli et al. (1996).}
\end{figure*}

We will briefly describe the first results from the analysis of
2D spectroscopy data on the internal kinematics of the objects under study.
One of the spectacular examples is NGC\,2950, an S0 galaxy, on the images of which
is clearly defined the secondary bar turned through approximately $60\degr$
with respect to the primary one (Fig. 1 and references in Table 1). The mismatch
between the position angle of the kinematic axis of
the velocity field of stars (PA$_{kin}$) and the location of the line of nodes
(defined from the outer isophotes) reaches $\rm \Delta PA=10-20\degr$ at distances
$r<8''$ from the centre. The kinematic  axis turns in the direction opposite
with respect to the line of nodes, as compared to the position angle of the
inner isophotes (Fig. 2). It is precisely this behaviour that is characteristic
for the triaxial potential of the bar (Monnet et al., 1992; Moiseev \&
Mustsevoi, 2000). It can be seen from Fig. 2 that outside the secondary bar
the kinematic axis intersects the line of nodes, and further at $r>10''$
its position angle keeps decreasing, which is associated now with the influence
of the primary bar.

\begin{figure*}
\centerline{\psfig{figure=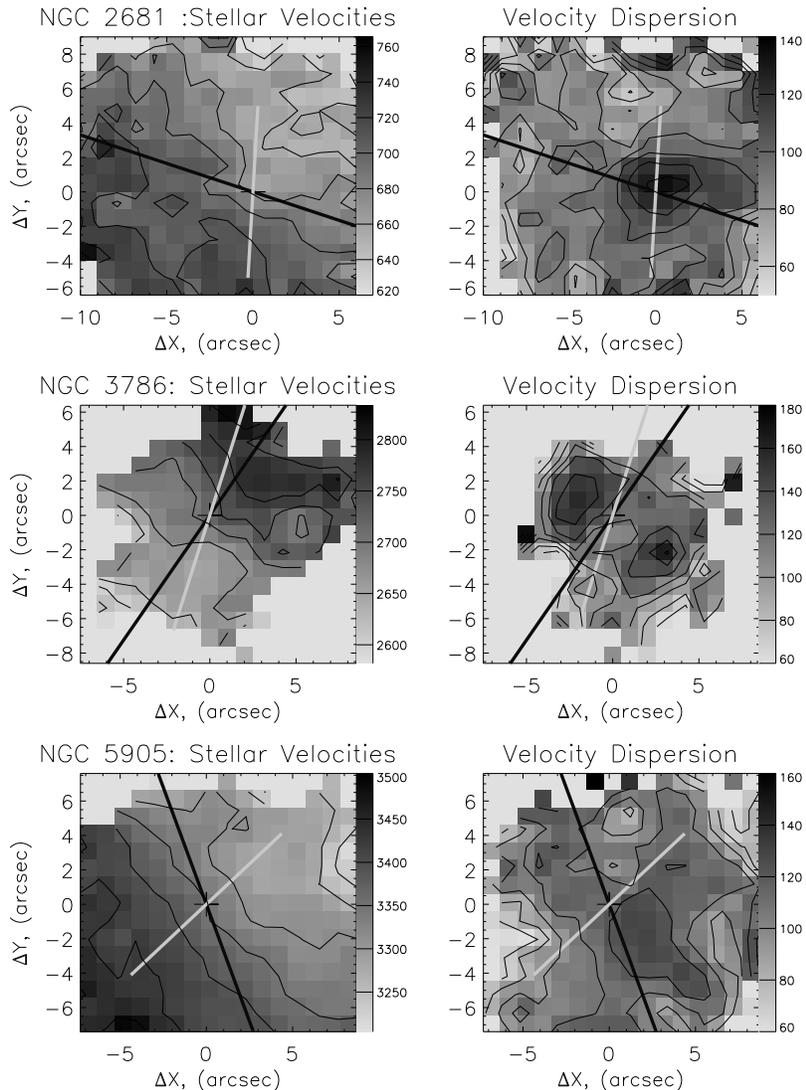,width=11cm}}
\caption{Radial velocity fields of stars (left) and maps of velocity dispersion of stars (right) in the
galaxies  NGC~2681, NGC~3780 and NGC~5905.
The scale in km/s. The solid black lines show the orientation of the outer
bar and the white lines show that of the inner one (from photometry data, see Table.~2).}
\end{figure*}

Thus, the secondary bar, which is seen on the images of NGC\,2950, shows
itself in the velocity field as well. It is interesting, however,
 though a central ellipsoidal structure
 of $r\sim 5''$ in size is seen on  the velocity dispersion map (Fig. 2), but
its major axis coincides with the outer (not with the inner) bar. Note that in
a galaxy consisting only of a disk and a bulge, the radial velocity
dispersion of stars must be symmetric along the radius, i.e. the map $\sigma_*$
has the pattern of concentric ellipses whose major axis is coincident with
the line of nodes of the disk. In the presence of a bar, as it is shown in
the papers dealing with numerical modeling (Miller \& Smith, 1979; Vauterin \&
Dejonghe, 1997), the distribution of spatial components of the radial velocity
 dispersion changes so that the character of symmetry on the map of the radial
velocity dispersion changes also. The distribution of $\sigma_*$ in the region of
the bar will be symmetric about the bar axis but not about the line of nodes
of the disk.

However, the observed radial velocity dispersion distributions of stars
in the galaxies explored are considerably more diversified (see Fig. 3).
Apart from the expected elliptical structures (NGC\,470, NGC\,2950, NGC\,2681)
also peaks of $\sigma_*$ are observed, which are displaced by a few arcseconds
with respect to the dynamic centre (NGC\,3368). These peaks have a more complex
shape (NGC\,5905). Two symmetric peaks are seen in the Seyfert galaxy NGC\,3786
 at a distance of $3-5''$ from the centre. The velocity dispersion gradient here
is about $\sim 50$ km/s. Unfortunately, in literature there are
absent any systematic observational data on the two-dimensional distributions
of the velocity dispersions in the bars. All the papers are generally
restricted to one or two long-slit cross-sections. The most consistent approach to
observational studies of asymmetry of the distribution of the radial velocity
dispersion in the bars is presented in the paper by Kormendy (1983) by the
example of NGC\,936. Emsellem et al. (2001) have found ``central drops'' in
the radial distributions of the radial velocity dispersion in several
double-barred  galaxies, but they did not explain them. Note, that if we
had studied the stellar kinematics in NGC\,3786 with a long-slit spectrograph,
we should have discovered the ``drops'' in the $\sigma_*$ distribution similar
to those described by Emsellem et al. (2001).

Analysis of the velocity fields of gas and stars has shown that in all the galaxies
the value of the PA$_{kin}$ in the circumnuclear region is different from the
position of the line of nodes by $10\degr-25\degr$, which is suggestive of
considerable non-circular motions. But only in a few galaxies (NGC\,2950, NGC\,3368 and
less reliably in NGC\,5850) the PA$_{kin}$ changes on the scales of the secondary
bar, and these changes being in opposite direction with respect to the PA of the inner
isophotes. This may point to the dynamic decoupling  of the inner bar. The
turn of the PA$_{kin}$ in the rest of the objects is likely to be related to
the influence of the outer bar.

In the circumnuclear regions of NGC\,3368, NGC\,3768 and NGC\,5905 the position
of the kinematic axis of the PA$_{kin}$ (stars), which is determined from the velocity field of stars, differs
systematically by $10\degr-20\degr$ from the PA$_{kin}$ (gas) defined from the
velocity fields of ionized gas. This feature is not related directly to the
secondary bar but is the reflection of the fact that gas and stars move in the
bar in different manner (see Section 1). The case of NGC\,3368 is of interest. Here
the radial relationship of the PA$_{kin}$ (stars) at $r=2-15''$ copies the
behaviour of the PA$_{kin}$ (gas), but it displaced systematically with
reference to the latter by $\sim 15\degr$ (the measurement accuracy of the
position angle being close to $1\degr$ everywhere (see Fig. 4). Note that,
using the results of numerical modeling, Shaw et al. (1993) predicted a similar
effect (``phase shift between gas and stars''), which must be observed in the
central regions of the barred galaxies  between two ILR resonances.

\begin{figure}
\centerline{\psfig{figure=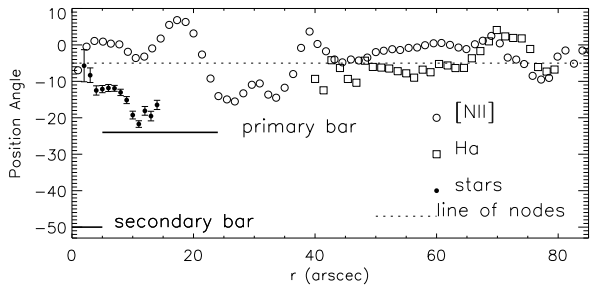,width=8.5 cm}}
\caption{Comparison of the kinematic axis orientation determined from the
velocity field of stars (filled circles) and from the lines of ionized gas
$\rm H_\alpha$ (squares) and [NII] (circles) in the galaxy  NGC\,3368.}
\end{figure}

\section{Conclusions}

Thus we made the first attempt to study the internal kinematics of double-barred
galaxies. In the central regions of all the investigated galaxies deviations
from the purely circular rotation are being detected (turn of the kinematic
axis in the velocity fields, different for stars and ionized gas, i.e.
asymmetry in the observed distribution of the radial velocity dispersion of
stars). However, the region of the inner bar seems to be dynamically decoupled
from the outer bar but in a few cases (NGC\,2950, NGC\,3368, NGC\,5850).
It is not improbable that this is related to the fact that the dynamically
independent secondary bar is a considerably rarer phenomenon than it follows
from the analysis of images of galaxies. This conclusion is consistent with
some theoretical models (Friedli \& Martinet, 1993); Khoperskov et al., 2001)
which suggest that the secondary bar is a relatively short-lived structure
inside the long-lived large-scale bar. Moreover, Khoperskov et al. (2001) show
that the three components of the velocity dispersion of stars in a
galaxy with a bar have different shape of the distribution in the disk plane.
The radial velocity dispersion $\sigma_*$ will have then a rather complex
distribution in the sky plane. The shape of distribution $\sigma_*$ is
first of all defined by the parameters of orientation of the bar and the disk
relative to the observer. Changing these parameters, one can obtain the
distribution $\sigma_*$ with a drop at the centre (like in NGC\,3786) or
elongated perpendicularly to the inner bar, like in NGC\,2950. We consider the
dynamical three-dimensional modeling of particular galaxies with the use of
all the kinematic data that we have acquired and with the involvement of the
available photometric data to be the next step in the study of double bars.
This is contemplated to be done in the next papers.

\begin{acknowledgements}
The author is grateful to V.L. Afanasiev for numerous methodological advises
and help in observations with the 6m telescope and to O.K. Sil'chenko for
continuous interest to the work.
This research has made use of the NASA/IPAC
Extragalactic Database (NED) which is operated by the Jet Propulsion Laboratory,
California Institute of Technology, under contract with the National Aeronautics and
Space Administration.
\end{acknowledgements}

\end{document}